# Coherence toroidal vortices and statistic-veiled correlation topologies

Keyu Zhou[1,2], Yaning Zhou[1,2], Ao Zhou[1,2], Zhao Zhang[1,2], Jinzhan Zhong[3], Houan Teng[4], Chunhao Liang[1,2,†], Qiwen Zhan[4,†], Yangjian Cai[1,2,†], Xin Liu[1,2,†]

[1]Shandong Provincial Key Laboratory of Light Field Manipulation Physics and Applications & School of Physics and Optoelectronics, Shandong Normal University, Jinan 250358, China

[2]Collaborative Innovation Center of Light Manipulations and Applications, Shandong Normal University, Jinan 250358, China

[3]School of Optical-Electrical and Computer Engineering, University of Shanghai for Science and Technology, Shanghai 200093, China

[4]Zhejiang Key Laboratory of 3D Micro/Nano Fabrication and Characterization, Department of Electronic and Information Engineering, School of Engineering, Westlake University, Hangzhou 310030, China

[†]Corresponding Authors: chunhaoliang@sdnu.edu.cn; zhanqiwen@westlake.edu.cn; yangjiancai@sdnu.edu.cn; xinliu@sdnu.edu.cn;

**Toroidal vortices in fluid and gas dynamics underpin a broad spectrum of scientific and technological fields, from elementary particle physics to condensed matter systems, and have recently garnered significant attention in optics because of their inherent topological stability. Here we report the experimental observation of toroidal vortices in stochastic optical wavefields with partial coherence, termed coherence toroidal vortices, which eliminates deterministic topological signatures in conventional optical degrees of freedom while unveiling statistically hidden correlation topologies. These underlying topologies—including both fundamental and higher-order hopfionic textures—emerge exclusively in second-order field correlations and are accessible only through statistical measurements. We further examine the impact of chaotic channels on the stability of these statistically veiled correlation topologies, demonstrating that their topological invariants remain robust under realistic environmental perturbations. These findings are experimentally validated and offer novel insights into the potential of toroidal light vortices serving as controllable channels for directional energy and information transfer within complex media.**

**Introduction**

Vortices are inherent to any wave phenomena, serving as fundamental carriers of the orbital angular momentum (OAM) associated with energy transport and evolution. Toroidal vortices, as new types of nontrivial topology, are intriguing three-dimensional (3D) torus-shaped structures characterized by the rotational motion localized around a closed-loop vortex core [1]. These toroidal topological patterns arise ubiquitously in both natural phenomena and fluid systems [2-8], manifest in various forms, spanning from smoke rings and cavitation bubbles in turbulent fluids [2,3], to thermally driven formations such as mushroom clouds [4], as well as biologically generated vortices involved in microscale propulsion [5], seed dispersal [6,7], and cardiovascular transport [8]. In optical community, toroidal vortices of light have been recently demonstrated to be realizable either as toris of electromagnetic fields [9-11] or as closed-loop spatiotemporal optical vortices [12-17]. While toroidal light vortices are known to occur in coherent waves but are ill defined in partially coherent stochastic waves where statistics are required to quantify the phase structure [18,19].

Nontrivial topological textures, such as knotted [20,21], skyrmionic and hopfionic [22-24] structures, have been observed across a broad range of physical systems, including liquid crystals [25-27], fluid dynamics and magnetism [28-30], as well as quantum fields [31,32]. Among these, 3D optical toroidal topologies, exemplified by optical skyrmions and hopfions [33,34], have attracted considerable interest due to their particle-like characteristics and their ability to exhibit a variety of topologically protected properties [35-39]. Consequently, optical realizations of these 3D toroidal topologies have been extensively pursued in topologically structured light, primarily through the controlled shaping of toroidal vortices in 3D space [40-46]. To date, however, these studies have been overwhelmingly confined to deterministic toroidal vortices in fully coherent optical fields [47]. By contrast, stochastic wavefields are generally considered incapable of supporting well-defined toroidal topologies, as vortex-carrying beams gain or lose OAM from an interaction with random media or chaotic channels [48].

Spatial correlations exist in many different physical systems, and the study of their origin and nature is one of the primary roles of statistical physics. In this work, we experimentally demonstrate the existence of toroidal vortices and their associated statistically veiled correlation topologies in stochastic optical waves, which emerge exclusively in second-order field correlations and are observed only through statistical evaluation. Because optical coherence measures the extent of statistical correlations of light fields at pairs of space-time points, we thereby use the term *coherence toroidal vortices* to refer to them. Coherence toroidal vortices reveal statistically veiled fundamental and higher-order 3D hopfionic topologies. Furthermore, we experimentally investigate the stability of these statistically veiled correlation topologies under atmospheric turbulence and demonstrate that their topological invariants are preserved across a chaotic medium, confirming their intrinsic robustness to environmental perturbations. Our findings establish a novel perspective for constructing statistically veiled toroidal light vortices in stochastic matter waves and highlight the unique advantages of their topological robustness for applications in realistic environments.

**Results and discussion**

**Concept and theory**

The construction of optical toroidal topologies typically relies on paraxial beam propagation and is realized by expanding the polynomial associated with a specific topology into a superposition of exact solutions to the paraxial wave equation. The resulting physical field can be expressed in terms of paraxially propagating Laguerre-Gaussian ($\text{LG}_p^\ell$) mode, which constitute a complete set of free-space solutions to the paraxial wave equation. Accordingly, the field is described by the following complex scalar function,

$$\text{LG}_p^\ell(r,\theta,z) = \frac{r^{|\ell|} e^{i\ell\theta}}{w^{|\ell|+1}(z)} L_p^{|\ell|}\left(\frac{2r^2}{w^2(z)}\right) \exp\left(-\frac{r^2}{w^2(z)}\right) \qquad (1)$$
$$\times \exp\left[-ik\, 2r^2/2R(z) - ikz - i(2p+|\ell|+1)\chi(z)\right],$$

characterized by the topological charge $\ell$ and the radial index $p$, where $w(z)$, $R(z)$ and $\chi(z)$ denote, respectively, the beam waist, radius of curvature of the wavefront and Gouy phase, and $L_p^{|\ell|}(\cdot)$ is an associated Laguerre polynomial, $k$ is the wavenumber. The fields described by Eq.

(1) are deterministic, exhibiting well-defined twist and edge phase singularities in the transverse plane, as shown in Fig. 1**a** and 1**b**. The superposition of such paraxially propagating modes, given by $E(\vec{r}, z) = \sum_{p,\ell} c_{p,\ell} \text{LG}_p^\ell(\vec{r}, z)$ and $\vec{r} = (x, y)$ is a 2D transverse coordinate, results in a deterministic toroidal phase topology throughout the 3D $(\vec{r}, z)$ field for appropriately chosen mode indices $(p, \ell)$ and weighting coefficients $c_{p,\ell}$ [49]. For partially coherent stochastic wavefields, a set of complex field amplitudes of the form $\text{LG}_p^\ell(\vec{r})\mathcal{R}(\vec{r})$ can be regarded as an extended quasi-monochromatic pseudothermal source [18,50], where $\mathcal{R}(\vec{r})$ denotes a random complex field characterized by a prescribed spatial correlation function [Methods]. Specifically, its second-order correlation satisfies $\langle \mathcal{R}(\vec{r}_1) \cdot \mathcal{R}^*(\vec{r}_2) \rangle = \exp[-|\vec{r}_1 - \vec{r}_2|^2/\sigma^2]$, where $\sigma$ is the transverse coherence length at $z = 0$ and $\langle \cdot \rangle$ denotes ensemble average. The instantaneous complex amplitude of the resulting stochastic waves exhibits a random spatial distribution (Fig. 1**c**) and therefore does not support the formation of paraxial toroidal vortices, in contrast to deterministic vortex beams. This is because the twist phase structure and edge dislocations required for such toroidal topologies are no longer preserved in the field at the first order. Instead, these topologic textures are transferred to higher-order statistical correlations of the field. As shown in Fig. 1**d**, the corresponding correlation function $G(\vec{r}_1, \vec{r}_2) = \langle E(\vec{r}_1)E^*(\vec{r}_2) \rangle$ reveals a stationary vortex structure, characterized by the emergence of twist and edge 'coherence' singularities [18]. Figure 1**e** further illustrates the evolution of instantaneous intensity cross-sections alongside the progressive emergence of toroidal structures in the correlation domain as the ensemble size increases. Notably, while individual realizations remain spatially random, the stochastic field retains the capacity to encode hopfion-like toroidal vortices within its higher-order correlations. We refer to this phenomenon as *statistically veiled correlation topologies*, emphasizing that the underlying topological structure is not observable in single-shot fields but becomes manifest only through statistical correlation.

**Experimental constructions and observation**

In the experiment, stochastic optical fields associated with a target coherence toroidal vortex are synthesized via coherent-mode decomposition [51]. Each realization of the stochastic waves is encoded as a phase-only computer-generated hologram (CGH) and implemented on a spatial light modulator (SLM) within a 4*f* spatial filtering system. A reference arm is split prior to the SLM to form a Mach-Zehnder interferometer, enabling retrieval of the complex field of the generated beam. The camera records the field at the beam waist plane; accordingly, the CGH encodes the corresponding field at $z = -L$, where $L$ denotes the total propagation distance from the SLM image plane to the camera. Details of experimental setup and light field reconstruction are given in the Methods. From the off-axis interference fringes, the instantaneous complex field of the stochastic optical fields $E(\vec{r}, 0)$ at a single transverse plane $(x, y, 0)$ is retrieved. The entire 3D field $E(\vec{r}, z)$ in $(x - y - z)$ space is subsequently recovered by numerically propagating the measured field using the angular spectrum method in both forward and backward directions [52]. Based on the ensemble of reconstructed fields, the second-order correlation function at a prescribed reference point is evaluated via statistical correlation, given by $G(\vec{r}_1, \vec{r}_2, z) = \langle E(\vec{r}_1, z) E^*(\vec{r}_2, z) \rangle$. To characterize the correlation features, one coordinate is fixed at $\vec{r}_2 = (0,0)$, thereby reducing the 5D correlation function to a 3D representation. Figures 2**a**,2**b** and 2**f**,2**g** present the complex amplitude profiles of the correlation function $G(\vec{r}, 0,0)$ in the toroidal plane, evaluated by ensemble averaging over 2500 speckle realizations. A clear phase dislocation is observed at the spatial location corresponding to the toroidal intensity null of the correlation function [Fig. 2**c**, 2**h**]. The correlation function evaluated in the poroidal plane reveals an twist coherence dislocation [see Supplementary Fig. S1], forming a characteristic toroidal vortex structure, as shown in Figs. 2**c** and 2**h**. Figure 2**d** and 2**i** show the extracted 3D iso-intensity surfaces reconstructed from Fig. 2**c** and 2**h**, using an isovalue of 3.5% of the peak intensity. These results indicate the formation of a coherence toroidal vortex with a clear vortex core. In addition, the statistically veiled toroidal correlation topologies are visualized in Figs. 2**e** and 2**j**, where colored curves mapped onto the toroidal surface represent the underlying 3D phase structure. Notably, these statistical-veiled topological trajectories remain unlinked and unknotted, owing to the absence of vortex components in the toroidal plane [12,15,44].

We have experimentally demonstrated the existence of toroidal light vortices within stochastic waves—termed coherence toroidal vortices—rather than in coherent (deterministic) continuous or polychromatic light fields [12]. Topologically, toroidal light vortices carrying longitudinal OAM can further form optical hopfions, which consist of an infinite series of tori layers interwoven by closed loops [15,44]. Accordingly, hopfion-like statistic-veiled correlation topologies can be generated from fundamental toroidal vortices with $\ell \neq 0$. Figures 3**a**,3**b** and 3**f**,3**g** show the amplitude and phase profiles of the correlation function $G(\vec{r}_1, \vec{r}_2)$ with $\vec{r}_2 = (0.15\text{mm}, 0.15\text{mm})$ in the toroidal plane at $z = 0$. Twist and edge phase dislocations are clearly observed at the spatial locations corresponding to the null core and the toroidal intensity null of the correlation function, respectively [Fig. 3**c**, 3**h** and Supplementary Fig. S2]. Notably, in the longitudinal component of $G(\vec{r}_1, \vec{r}_2, z)$ [Fig. 3**c** and 3**h**], the amplitude distribution preserves rotational symmetry, while the phase distribution evolves with the observation plane due to the toroidal vortex structure ($\ell = 1$). The intensity isosurfaces in Figs. 3**d** and 3**i** further confirm the presence of coherence toroidal vortices in the correlation function of the stochastic waves. The hopfionic correlation topologies extracted from Figs. 3**d** and 3**i**, as visualized in Figs. 3**e** and 3**j**, wind around both the poloidal and the toroidal vortex cores with a crossing number of $2\ell = 2$, thereby forming a statistically veiled optical hopfion.

Linked loops residing on the surface of a complete torus constitute a defining feature of a scalar optical hopfion, wherein the number of crossings between any two linked equiphase lines is given by $2\ell$. By introducing higher-order longitudinal OAM ($\ell \geq 2$) into the coherence toroidal vortices constructed above, higher-order statistically veiled optical hopfions can be generated. Figure 4 presents the experimental realization of high-order statistically veiled optical hopfions extracted from the coherence toroidal vortex for (a) $\ell = 2$ and (b) $\ell = 3$. The representative equiphase lines are linked, with crossing numbers of 4 and 6, respectively, in agreement with $2\ell$, thereby clearly revealing the topological features of statistically veiled hopfions.

**Topological robustness of statistic-veiled correlation topologies**

Notably, the coherence toroidal vortices and the associated statistically veiled topologies reported to date are encoded in the second-order field correlations of stochastic waves, rather than in deterministic optical degrees of freedom. The latter are inherently susceptible to external perturbations and tend to degrade in distorting channels, leading to the progressive loss of phase-defined topological features in the presence of noise [see Supplementary Fig. S3]. In contrast, correlation-encoded topologies exhibit an intrinsic resilience to such disturbances. Building on this distinction, we further demonstrate that these statistically veiled correlation topologies can be harnessed to achieve enhanced robustness against propagation through chaotic channels. To experimentally test this effect, we emulate dynamic atmospheric turbulence with a hotplate operating at different temperatures, thereby spanning a range of perturbation strengths as a representative real-world propagation environment [Methods]. The experimental results for the constructed statistically veiled correlation topologies at different temperatures are presented in Fig. 5. The results show that the correlation topologies can be accurately reconstructed from the corresponding coherence toroidal vortices in the presence of atmospheric turbulence, while the number of crossings remains invariant ($N = 2\ell$) across all turbulence strengths [Supplementary Fig. S4].

**Conclusions**

In conclusion, we theoretically and experimentally establish a continuous family of toroidal vortices—termed coherence toroidal vortices of light—and reveal the emergence of statistically veiled correlation topologies in stochastic wavefields. In contrast to conventional deterministic topology, both fundamental and higher-order hopfionic textures arise here in the second-order field correlations rather than in the instantaneous field itself, highlighting a fundamentally distinct topological paradigm. The veiled topology, encoded in the equiphase-line structure of the correlation function, becomes accessible only through statistical correlations and is governed uniquely by the longitudinal OAM of the coherence toroidal vortices. Moreover, compared with classically defined deterministic topologies, the statistically veiled topology introduced here exhibits enhanced robustness against perturbations, enabling the preservation of topological invariants under disorder and thereby offering practical advantages for applications in complex or noisy environments.

**Methods**

**Experimental setup for generating and constructing coherence toroidal vortices**

In the experiment, we employed a Mach-Zehnder interferometer analogous to that in [53] to generate and characterize coherence toroidal vortices and the associated statistically veiled topologies [Supplementary Note 3], as illustrated in Fig. 6. In this configuration, a heating plate is positioned at the output plane of the $4f$ system to emulate a dynamically varying refractive-index distribution within a chaotic channel, with the level of randomness controlled via temperature modulation. The optical axis is aligned parallel to, and maintained at a separation of approximately 2 cm from, the surface of the heating plate, resulting in an effective propagation length of 300 mm through the perturbed region. The SLM displays a sequence of phase-only CGH, computed from the complex amplitude of the target stochastic fields using a cosine-grating algorithm, thereby shaping the incident beam into the desired stochastic wavefields at the recording plane. The random complex field $\mathcal{R}(\vec{r})$ is calculated by $\mathcal{R}(\vec{r}) = \iint \sqrt{p(\vec{v})}\exp(-i2\pi\vec{r}\cdot\vec{v})d^2\vec{v}$, where $p(\vec{v})$ is the power spectral density of the resulting stochastic wavefields [Supplementary Note 3]. In the experiment, the transverse coherence length and beam waist are set to 1 mm and 0.5 mm, respectively. The complex amplitude of each instantaneous realization of the stochastic wavefield at a given plane is retrieved using the Mach-Zehnder interferometer. The 3D field distribution can be further reconstructed by numerically propagating the measured field both forward and backward using the angular spectrum method. Subsequently, a statistical ensemble comprising 2500 instantaneous realizations is employed to reconstruct and analyze the emergent 3D coherence toroidal vortex. The corresponding statistically veiled correlation topologies are then extracted from the reconstructed coherence vortex via an equiphase-line extraction method [Supplementary Note 4].

**Funding.** National Key Research and Development Program of China (2022YFA1404800 [Y.C.]), National Natural Science Foundation of China (12192254 [Y.C.], W2441005 [Y.C.], 12534014 [Y.C.], 12434012 [Q.Z.], 62535013 [Q.Z.], 12547149 [X.L.], 12374311 [C.L.], 12304367 [J.Z.]), Natural Science Foundation of Shandong Province (ZR2025ZD21 [Y.C.], ZR2023YQ006 [C.L.]), Key Research and Development Program of Shandong Province (2024JMRH0105 [Y.C.]), Taishan Scholar Project of Shandong Province (tsqn202312163 [C.L.]).

**Disclosures.** The authors declare no competing interests.

**Data availability.** The data that support the findings of this study are available from the corresponding authors upon reasonable request.

**Author contributions.** X.L. conceived the research and led the manuscript writing, K.Z., A.Z., and Z.Z. conducted the simulations and experiments. K.Z., Y.Z., and X.L. contributed to visualization. K.Z., Y.Z., J.Z., H.T., C.L., and X.L. performed the data analysis. C.L., Q.Z., Y.C., and X.L. supervised the project. All authors contributed to the discussion and revision of the manuscript.

**Figures and legends**

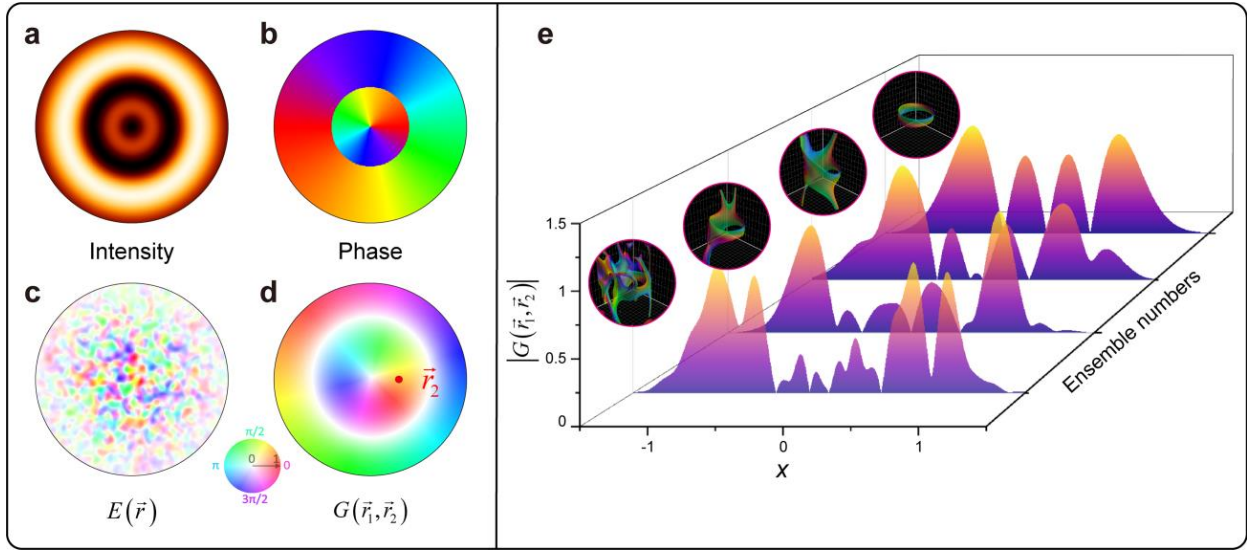

**Fig. 1 | Comparisons of waveforms between deterministic and stochastic vortex-carrying light beams. a.** Amplitude and **b.** phase distributions of a deterministic LG beam with topological charge $\ell = 1$ and radial index $p = 1$. **c.** Complex amplitude of stochastic LG beams ($\ell = 1$, $p = 1$), exhibiting a random transverse distribution in the instantaneous field. **d.** Complex amplitude of the statistical correlation derived from the stochastic LG beams in **c**, revealing a stationary vortex structure (twist and edge 'coherence singularity'). **e.** Evolution of instantaneous intensity cross-sections and the corresponding emergence of coherence toroidal vortices with increasing ensemble size. The brightness and hue denote magnitude and phase distributions respectively.

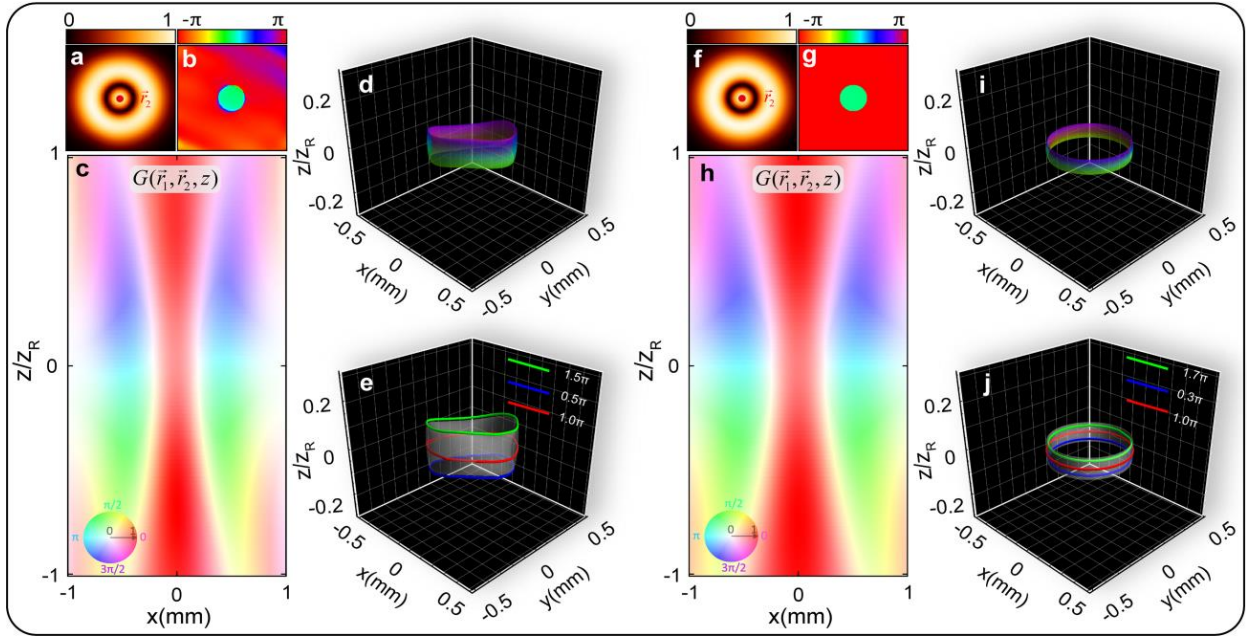

**Fig. 2 | Experimental (left) and simulated (right) generation of a fundamental coherence toroidal vortex of $\ell = 0$ and statistic-veiled correlation topologies.** (**a**,**f**) Amplitude and (**b**,**g**) phase distributions of the second-order correlation function $G(\vec{r}_1, \vec{r}_2) = \langle E(\vec{r}_1, 0)E^*(\vec{r}_2, 0)\rangle$ of the stochastic waves, $E(\vec{r}, 0) = \mathrm{LG}_0^0 + 2\mathrm{LG}_1^0$, in the transverse $(x - y)$ (toroidal) plane. The red dot marks the reference point $\vec{r}_2 = (0,0)$ used in the statistical correlation evaluation. (**c**,**h**) Complex field profiles of the correlation function $G(\vec{r}_1, \vec{r}_2, z)$ in the longitudinal $(x - z)$ (poloidal) plane. Longitudinal coherence singularities indicate the locations of the coherence toroidal vortices. Brightness and color encode the magnitude and phase, respectively. (**d**,**i**) 3D intensity isosurfaces and corresponding phase mapping of the coherence toroidal vortices, reconstructed from (**c**,**h**), with an isovalue of 3.5%. (**e**,**j**) Extracted statistic-veiled toroidal correlation topologies, featured by 3D colored equiphase curves on the torus surface of the coherence toroidal vortices.

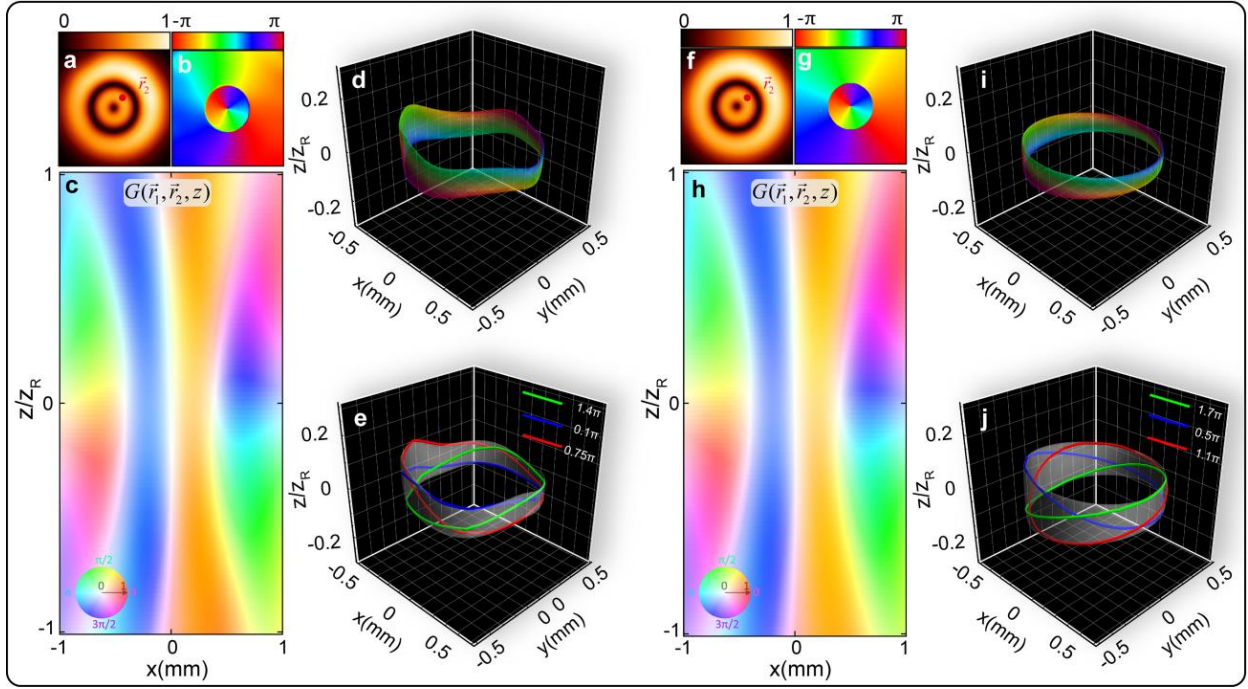

**Fig. 3 | Experimental (left) and simulated (right) generation of a coherence hopfionic toroidal vortex of $\ell = 1$ and statistic-veiled correlation topologies.** (**a,f**) Amplitude and (**b,g**) phase distributions of the second-order correlation function $G(\vec{r}_1, \vec{r}_2) = \langle E(\vec{r}_1, 0)E^*(\vec{r}_2, 0)\rangle$ of the stochastic waves, $E(\vec{r}, 0) = \text{LG}_0^1 + 2\text{LG}_1^1$, in the transverse $(x-y)$ (toroidal) plane. The red dot marks the reference point $\vec{r}_2$ used in the statistical correlation evaluation. (**c,h**) Complex field profiles of the correlation function $G(\vec{r}_1, \vec{r}_2, z)$ in the longitudinal $(x-z)$ (poloidal) plane. Longitudinal coherence singularities indicate the locations of the coherence toroidal vortices. Brightness and color encode the magnitude and phase, respectively. (**d,i**) 3D intensity isosurfaces and corresponding phase mapping of the coherence toroidal vortices, reconstructed from (**c,h**), with an isovalue of 3.5%. (**e,j**) Extracted statistic-veiled toroidal correlation hopfion-like topologies, featured by 3D colored equiphase curves on the torus surface of the coherence toroidal vortices.

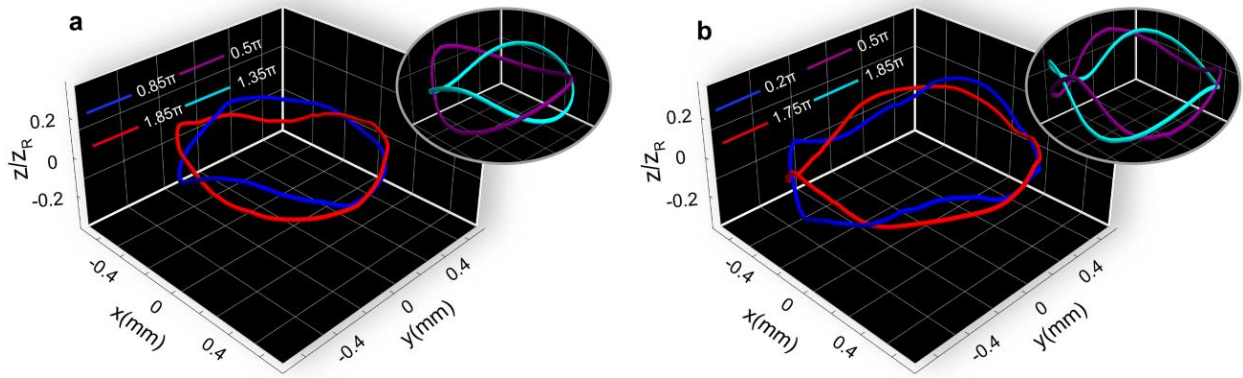

**Fig. 4 | Experimental observation of higher-order statistically veiled optical hopfions, represented by two equiphase lines with crossing numbers of (a) $2\ell = 4$ and (b) $2\ell = 6$.** Insets show the corresponding theoretical results.

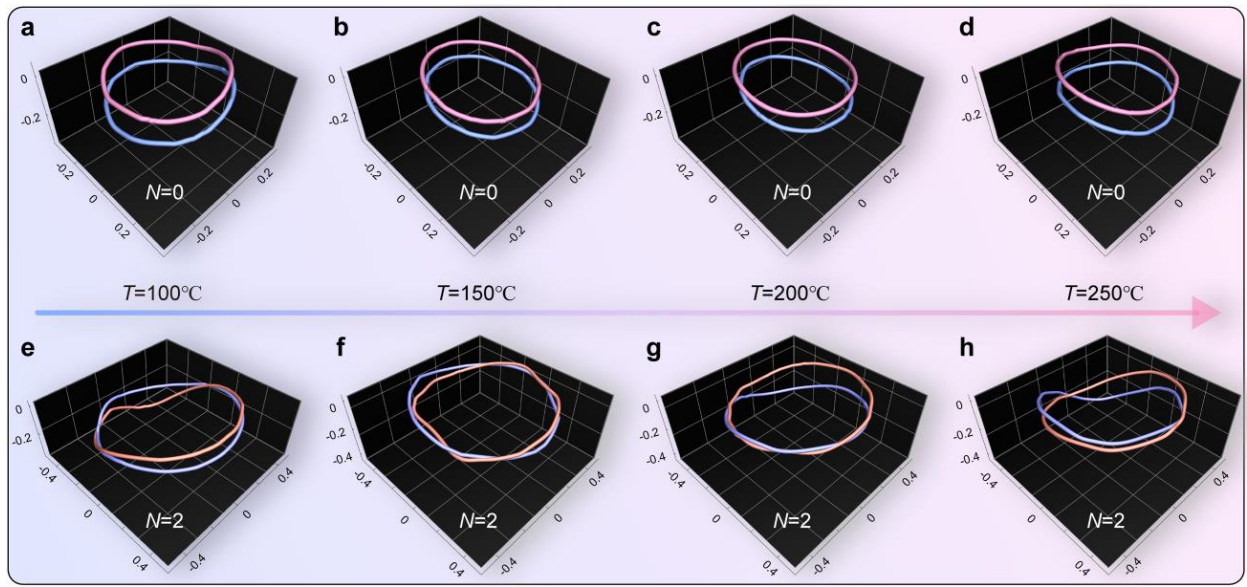

**Fig. 5 | Statistically veiled correlation topologies extracted from (a-d) fundamental and (e-h) coherence hopfionic toroidal vortices under atmospheric turbulence at different temperatures.** All results are obtained via ensemble averaging over 2500 speckle realizations for different topologies, demonstrating that the crossing number $N$ remains a topological invariant in the presence of turbulence [see Supplementary Fig. S4].

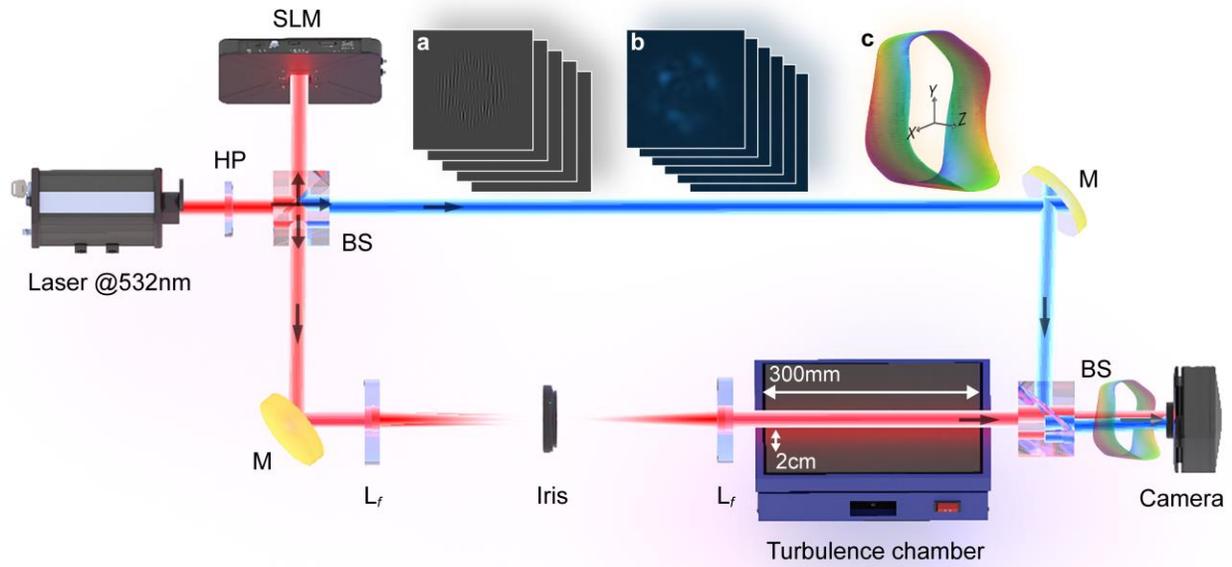

**Fig. 6 | Experimental setup. a.** Holograms used to synthesize coherence toroidal vortices with target statistically veiled correlation topologies. **b.** Experimentally measured interference fringes of the instantaneous fields of the resulting stochastic wavefields. **c.** Experimentally constructed coherence toroidal vortex, characterized by a 3D torus iso-intensity profile with its phase structure encoded in the second-order field correlation $G(\vec{r}_1, \vec{r}_2)$. HP, half-wave plate; SLM, spatial light modulator; $L_f$, lens with a focal length of $f$=200 mm; M, Mirror.